\documentclass[
showkeys,
superscriptaddress,
amsmath,amssymb,
reprint,
prl,
aps]{revtex4-1}

\usepackage{graphicx}
\usepackage{hyperref}

\newcommand{\beginsupplement}{%
        \setcounter{table}{0}
        \renewcommand{\thetable}{S\arabic{table}}
        \setcounter{figure}{0}
        \renewcommand{\thefigure}{S\arabic{figure}}
        \setcounter{equation}{0}
        \renewcommand{\theequation}{S\arabic{equation}}%
}

\begin{document}

\title{A Method of Determining Excited-States for Quantum Computation}

\author{Pejman Jouzdani}
\email{Corresponding author, email: jouzdanip@fusion.gat.com}
\affiliation{ Oak Ridge Associated Universities, Oak Ridge, TN}
\author{Stefan Bringuier}
\thanks{Principal investigator}
%\email{bringuiers@fusion.gat.com}
\affiliation{ General Atomics, San Diego, CA}
\author{Mark Kostuk}
%\email{kostukm@fusion.gat.com}
\affiliation{ General Atomics, San Diego, CA}
\date{\today}

%%%%%%%%%%%%%%%%%%%%%%%%%%%%%%%%%%%%%%%%%%%%%%%%
%%%%%%%%%%%%%%%%%%%%%%%%%%%%%%%%%%%%%%%%%%%%%%%%
%%%%%%%%%%%%%%%%%%%%%%%%%%%%%%%%%%%%%%%%%%%%%%%%
%%%%%%%%%%%%%%%%%%%%%%%%%%%%%%%%%%%%%%%%%%%%%%%%
%%%%%%%%%%%%%%%%%%%%%%%%%%%%%%%%%%%%%%%%%%%%%%%%
%%%%%%%%%%%%%%%%%%
%%%%%%%%%%%%%%%%%%
%%%%%%%%%%%%%%%%%%      Abstract
%%%%%%%%%%%%%%%%%%
%%%%%%%%%%%%%%%%%%%%%%%%%%%%%%%%%%%%%%%%%%%%%%%%
%%%%%%%%%%%%%%%%%%%%%%%%%%%%%%%%%%%%%%%%%%%%%%%%
%%%%%%%%%%%%%%%%%%%%%%%%%%%%%%%%%%%%%%%%%%%%%%%%
%%%%%%%%%%%%%%%%%%%%%%%%%%%%%%%%%%%%%%%%%%%%%%%%
%%%%%%%%%%%%%%%%%%%%%%%%%%%%%%%%%%%%%%%%%%%%%%%%

\begin{abstract}
A method is presented in which the ground-state subspace is projected
out of a Hamiltonian representation. As a result of this projection, an effective Hamiltonian is constructed where its ground-state coincides with an excited-state of the original problem. Thus, low-lying excited-state energies can be calculated using existing hybrid-quantum classical techniques and variational algorithm(s) for determining  ground-state. The method is shown to be fully valid for the H$_2$ molecule. In addition, conditions for the method's success are discussed in terms of classes of Hamiltonians.
\end{abstract}

%\keywords{Hybrid quantum-classical algorithms; Quantum computing; Excited-state Chemistry}

\maketitle

%%%%%%%%%%%%%%%%%%%%%%%%%%%%%%%%%%%%%%%%%%%%%%%%
%%%%%%%%%%%%%%%%%%%%%%%%%%%%%%%%%%%%%%%%%%%%%%%%
%%%%%%%%%%%%%%%%%%%%%%%%%%%%%%%%%%%%%%%%%%%%%%%%
%%%%%%%%%%%%%%%%%%%%%%%%%%%%%%%%%%%%%%%%%%%%%%%%
%%%%%%%%%%%%%%%%%%%%%%%%%%%%%%%%%%%%%%%%%%%%%%%%
%%%%%%%%%%%%%%%%%%
%%%%%%%%%%%%%%%%%%
%%%%%%%%%%%%%%%%%%      Intro
%%%%%%%%%%%%%%%%%%
%%%%%%%%%%%%%%%%%%%%%%%%%%%%%%%%%%%%%%%%%%%%%%%%
%%%%%%%%%%%%%%%%%%%%%%%%%%%%%%%%%%%%%%%%%%%%%%%%
%%%%%%%%%%%%%%%%%%%%%%%%%%%%%%%%%%%%%%%%%%%%%%%%
%%%%%%%%%%%%%%%%%%%%%%%%%%%%%%%%%%%%%%%%%%%%%%%%
%%%%%%%%%%%%%%%%%%%%%%%%%%%%%%%%%%%%%%%%%%%%%%%%

The calculation of molecular ground-state and excited-state energies, or more generally the energy spectra of chemical and material systems, is an application of great interest in a gate-based quantum computational model. Some experimental success has been achieved by calculating the ground-state energies of small molecules and correlated materials \cite{AK17,BB16}. These investigations have benefited from research on hybrid quantum-classical (HQC) approach \cite{JRM16} and noisy intermediate-scale quantum (NISQ) hardware \cite{JP18}.\par

Briefly, the formulation of a quantum chemistry problem into a quantum computational model  is to state the problem Hamiltonian in the second quantized language, using  Jordan-Wigner or Bravyi-Kitaev transformation of fermionic operators, and re-casting the problem  in terms of Pauli operators \cite{BK02,BW35}. In the NISQ era, extraction of the energy spectra is difficult to achieve using methods like the phase estimation algorithm, due to the resource requirements. \par

Alternatively, a process known as the variational quantum eigensolver (VQE) has been experimentally demonstrated to provide the ground-state energy \cite{AP14}. This algorithm takes an ansatz composed of parametric unitary gates on quantum hardware and finds the minimum energy of the Hamiltonian using optimization on a classical computer. In quantum chemistry, the unitary coupled cluster (UCC) is a commonly employed class of ans\"atze \cite{RB89,JR18}. We point out that a key assumption in variational approaches within HQC computation is that the number of terms in the Hamiltonian is polynomial  with regards to the number of qubits \cite{AP14}.\par

Regarding excited-states, quantum subspace expansion (QSE) has been proposed \cite{JIC18} where a set of basis, constructed from the optimized ansatz, is used to represent an approximation to the problem Hamiltonian. This matrix is then diagonalized on a classical computer to extract excited-states. A generalized QSE is used to find the absorption spectrum \cite{RP19}. Other recent proposals are demonstrated to be successful for the H$_2$ molecule \cite{OH19}, and for the random transverse Ising model \cite{KN18}. Further strategies to truncate the number of qubits have also been introduced \cite{SB17}. \par

In this letter we propose a phenomenological approach to the calculation of low-lying excited-states of a given problem Hamiltonian. This is done by projecting out the ground-state subspace and constructing a \emph{projected Hamiltonian}. The projected Hamiltonian has a ground-state that coincides with the first excited-state of the problem Hamiltonian. Furthermore, the projected Hamiltonian is reduced to an \emph{effective Hamiltonian} by means of a set of considerations -- what we refer to as a \emph{covariance assertion}. This assertion  ensures that the number of terms in the effective Hamiltonian remains approximately the same as the number of terms used in the calculation of the ground-state energy of the original problem Hamiltonian. As long as the covariance assertion holds, this method can be used to iteratively extract eigenstate and eigenenergies. \par

%%%%%%%%%%%%%%%%%%%%%%%%%%%%%%%%%%%%%%%%%%%%%%%%
%%%%%%%%%%%%%%%%%%%%%%%%%%%%%%%%%%%%%%%%%%%%%%%%
%%%%%%%%%%%%%%%%%%%%%%%%%%%%%%%%%%%%%%%%%%%%%%%%
%%%%%%%%%%%%%%%%%%%%%%%%%%%%%%%%%%%%%%%%%%%%%%%%
%%%%%%%%%%%%%%%%%%%%%%%%%%%%%%%%%%%%%%%%%%%%%%%%
%%%%%%%%%%%%%%%%%%
%%%%%%%%%%%%%%%%%%
%%%%%%%%%%%%%%%%%%      Eq. 1
%%%%%%%%%%%%%%%%%%
%%%%%%%%%%%%%%%%%%%%%%%%%%%%%%%%%%%%%%%%%%%%%%%%
%%%%%%%%%%%%%%%%%%%%%%%%%%%%%%%%%%%%%%%%%%%%%%%%
%%%%%%%%%%%%%%%%%%%%%%%%%%%%%%%%%%%%%%%%%%%%%%%%
%%%%%%%%%%%%%%%%%%%%%%%%%%%%%%%%%%%%%%%%%%%%%%%%
%%%%%%%%%%%%%%%%%%%%%%%%%%%%%%%%%%%%%%%%%%%%%%%%

\emph{Model.---} The Hamiltonian explored in this letter has the general form
\begin{eqnarray}
\hat{H} = \sum_j^{N_h} \lambda_j \hat{h}_j,
\label{eq-1}
\end{eqnarray}
with $\hat{h}_j = \hat{O}^{(j)}_1 \otimes \cdots \otimes \hat{O}^{(j)}_N $, where $\otimes$ indicates the tensor-product, $N$ is the number of qubits, and $\lambda_j\in \mathbb{R}$. The $\hat{O}^{(j)}_q$ operator is any of the identity or Pauli matrices: $\{ \hat{I}, \hat{X}, \hat{Y}, \hat{Z}\}$ acting on the $q$-th qubit and belonging to the $j$-th term in the sum of Eq. (\ref{eq-1}). In addition, an operator such as $\hat{h}_j$ (a tensor-product of single-qubit operators) is referred to as a \emph{string operator}. In Eq. (\ref{eq-1}) every two string operators $\hat{h}_j$ and $\hat{h}_k$, $k\ne j$, differ by at least one $\hat{O}$, and   $N_h =\vert  \{\hat{h}_j\} \vert $ is the cardinality of the set $\{ \hat{h}_j\}$. \par

To illustrate our method, consider that a problem Hamiltonian  $\hat{H}^{[0]}$ in the form of Eq. (\ref{eq-1}) is provided where we added the superscript $[i]$ to refer to the $i$-th iterative step. In a HQC computational model, a variational  procedure
such as the VQE algorithm is performed, and one ground-state (of potentially many orthogonal degenerate ground-states) of the Hamiltonian is approximated. This ground-state solution is expressed in terms of a set of optimized hardware parameters $\{\alpha^*\}$ and a corresponding quantum circuit; that is,  a unitary operation $\hat{U}^{[0]}\equiv\hat{U}^{[0]}(\{\alpha^*\})$.  This ground-state can  be reproduced by applying $\hat{U}^{[0]}$ on the same initial state of the qubits used in the VQE process, usually  $\vert \phi_{int} \rangle = \vert 0\cdots 0 \rangle$.

%%%%%%%%%%%%%%%%%%%%%%%%%%%%%%%%%%%%%%%%%%%%%%%%
%%%%%%%%%%%%%%%%%%%%%%%%%%%%%%%%%%%%%%%%%%%%%%%%
%%%%%%%%%%%%%%%%%%%%%%%%%%%%%%%%%%%%%%%%%%%%%%%%
%%%%%%%%%%%%%%%%%%%%%%%%%%%%%%%%%%%%%%%%%%%%%%%%
%%%%%%%%%%%%%%%%%%%%%%%%%%%%%%%%%%%%%%%%%%%%%%%%
%%%%%%%%%%%%%%%%%%
%%%%%%%%%%%%%%%%%%
%%%%%%%%%%%%%%%%%%      Eq 2, 3
%%%%%%%%%%%%%%%%%%
%%%%%%%%%%%%%%%%%%%%%%%%%%%%%%%%%%%%%%%%%%%%%%%%
%%%%%%%%%%%%%%%%%%%%%%%%%%%%%%%%%%%%%%%%%%%%%%%%
%%%%%%%%%%%%%%%%%%%%%%%%%%%%%%%%%%%%%%%%%%%%%%%%
%%%%%%%%%%%%%%%%%%%%%%%%%%%%%%%%%%%%%%%%%%%%%%%%
%%%%%%%%%%%%%%%%%%%%%%%%%%%%%%%%%%%%%%%%%%%%%%%%

Now, we define a \emph{projection operator} $\hat{P}^{[0]}$, which is given by
\begin{eqnarray}
\hat{P}^{[0]}&=&
\hat{U}^{[0]} \,
\vert \phi_{int} \rangle
\langle \phi_{int} \vert \, (\hat{U}^{[0]})^{\dagger}.
\label{eq-2}
\end{eqnarray}
Therefore, the projection of $\hat{H}^{[0]}$ onto the complement of $\hat{P}^{[0]}$ is
\begin{eqnarray}
\hat{H}^{[1]} &=& (\hat{I} - \hat{P}^{[0]} ) \hat{H}^{[0]} (\hat{I}-\hat{P}^{[0]}) \nonumber \\
&=& \hat{H}^{[0]} - E^{[0]}_g \, \hat{P}^{[0]},
\label{eq-3}
\end{eqnarray}
where $E^{[0]}_g$ is the ground-state energy of $\hat{H}^{[0]}$. At this stage, application of VQE on $\hat{H}^{[1]}$ should result in the minimum energy of $\hat{H}^{[1]}$ which, for the non-degenerate case, coincides with the first excited-state of $\hat{H}^{[0]}$ \cite{SupplMat0}. \par

The protocol outlined above can be applied recursively by replacing $\hat{H}^{[0]}$ with $\hat{H}^{[1]}$ and repeating. In the case of a degenerate subspace, the obtained ground-state corresponds to a degenerate ground-state of the previous iteration. Thus, if the protocol fails to discover an excited-state, it will produce an \emph{orthogonal} degenerate ground-state of the Hamiltonian of the previous iteration. \par

One of the challenges in the iterative process above is handling the sum of terms in the projection operator $\hat{P}^{[0]}$, which in the general form is
\begin{eqnarray}
\hat{P}  =  \sum_{p} f_p \hat{S}_p,
\label{eq-4}
\end{eqnarray}
where $\hat{S}_p = \hat{O}^{(p)}_1 \otimes \cdots \otimes \hat{O}^{(p)}_N $, with $\hat{O} \in \{ \hat{I}, \hat{X}, \hat{Y}, \hat{Z}\}$, are string operators.
A coefficient $f_p$ is a real number since, from Eq.~(\ref{eq-2}), $\hat{P}$ is hermitian; thus, coefficients $\{f_p\}$ must be real. Herein lies the challenge, namely, that the number of the terms in the sum is not trivial. This means the Hamiltonian at every iteration does not necessarily maintain the same set of  string operators,  $\{ \hat{h}_j \}$, and perhaps the number $N_h$ grows exponentially. \par

Our simplification to the number of string operators scaling problem is what we characterize as the \emph{covariance assertion}: Only the string operators in $\hat{P}$ that are already in the Hamiltonian $\hat{H}$ of the previous iteration are retained and the rest discarded. Therefore, at all levels of iteration, the form of the Hamiltonian remains \emph{covariant}, in the sense that the problem Hamiltonians between iterative steps transform as
%%%%%%%%%%%%%%%%%%%%%%%%%%%%%%%%%%%%%%%%%%%%%%%%
%%%%%%%%%%%%%%%%%%%%%%%%%%%%%%%%%%%%%%%%%%%%%%%%
%%%%%%%%%%%%%%%%%%%%%%%%%%%%%%%%%%%%%%%%%%%%%%%%
%%%%%%%%%%%%%%%%%%%%%%%%%%%%%%%%%%%%%%%%%%%%%%%%
%%%%%%%%%%%%%%%%%%%%%%%%%%%%%%%%%%%%%%%%%%%%%%%%
%%%%%%%%%%%%%%%%%%
%%%%%%%%%%%%%%%%%%
%%%%%%%%%%%%%%%%%%      Eq 5-7
%%%%%%%%%%%%%%%%%%
%%%%%%%%%%%%%%%%%%%%%%%%%%%%%%%%%%%%%%%%%%%%%%%%
%%%%%%%%%%%%%%%%%%%%%%%%%%%%%%%%%%%%%%%%%%%%%%%%
%%%%%%%%%%%%%%%%%%%%%%%%%%%%%%%%%%%%%%%%%%%%%%%%
%%%%%%%%%%%%%%%%%%%%%%%%%%%%%%%%%%%%%%%%%%%%%%%%
%%%%%%%%%%%%%%%%%%%%%%%%%%%%%%%%%%%%%%%%%%%%%%%%
\begin{eqnarray}
\hat{H}^{[i]} = \sum_j^{N_h} \lambda^{[i]}_j \hat{h}_j \rightarrow \hat{H}^{[i+1]} = \sum_j^{N_h} \lambda^{[i+1]}_j \hat{h}_j,
\label{eq-5}
\end{eqnarray}
with renormalized coefficients
\begin{eqnarray}
\lambda^{[i+1]}_j  = \lambda_j^{[i]} - E_g^{[i]}\,\,f^{[i]}_j,
\label{eq-renorm}
\end{eqnarray}
from Eqs. (\ref{eq-1}), (\ref{eq-3}) and  (\ref{eq-4}). \par

The coefficients $f^{[i]}_j$ corresponding to the string operator $\hat{h}_j$ in Eq.~(\ref{eq-renorm}) are given by
\begin{eqnarray}
f^{[i]}_j &=& {\mathbf{Tr}}\left[ P^{[i]} \,
\hat{h}_j \right]  \nonumber \\
&=& \langle \phi_{int} \vert \,(\hat{U}^{[i]})^\dagger \,\hat{h}_j\, \hat{U}^{[i]} \,\vert \phi_{int} \rangle,
\label{eq-7}
\end{eqnarray}
where $\mathbf{Tr}$ is the trace. Within a HQC computational model, the second line in Eq.~(\ref{eq-7}) is  the output of the measurement of the $\hat{h}_j$ operator performed by the quantum device and VQE algorithm at the $i$-th iterative step.

From an experimental perspective, at a given iteration, beside the efforts to obtain the ground-state, there are no additional computations nor measurements needed to construct the effective Hamiltonian of the next iteration. \par

\begin{figure}[ht]
    \centering
    \includegraphics[scale=1]{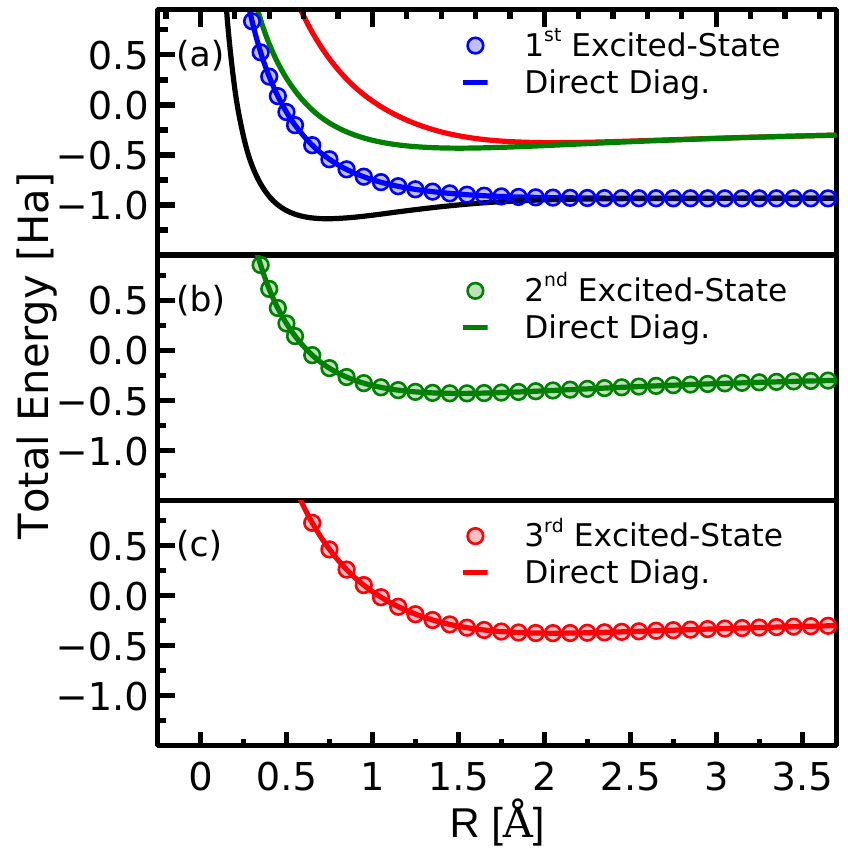}
    \caption{(Color online) Molecular hydrogen (H$_2$) binding curves. In (a) the results for excited-states using  direct diagonalization are shown as solid lines. In (a-c) the circles are the excited-state obtained by using the iterative method outlined in this letter. The total energy includes the nuclear-nuclear repulsion energy.}
    \label{fig-fig0}
\end{figure}

We now further motivate this method with the numerical (classically computed) results of the hydrogen molecular binding-energy curve including excited-states obtained using this method. The two-qubit Hamiltonian of this problem is given by
%%%%%%%%%%%%%%%%%%%%%%%%%%%%%%%%%%%%%%%%%%%%%%%%
%%%%%%%%%%%%%%%%%%%%%%%%%%%%%%%%%%%%%%%%%%%%%%%%
%%%%%%%%%%%%%%%%%%%%%%%%%%%%%%%%%%%%%%%%%%%%%%%%
%%%%%%%%%%%%%%%%%%%%%%%%%%%%%%%%%%%%%%%%%%%%%%%%
%%%%%%%%%%%%%%%%%%%%%%%%%%%%%%%%%%%%%%%%%%%%%%%%
%%%%%%%%%%%%%%%%%%
%%%%%%%%%%%%%%%%%%
%%%%%%%%%%%%%%%%%%      Eq 8
%%%%%%%%%%%%%%%%%%
%%%%%%%%%%%%%%%%%%%%%%%%%%%%%%%%%%%%%%%%%%%%%%%%
%%%%%%%%%%%%%%%%%%%%%%%%%%%%%%%%%%%%%%%%%%%%%%%%
%%%%%%%%%%%%%%%%%%%%%%%%%%%%%%%%%%%%%%%%%%%%%%%%
%%%%%%%%%%%%%%%%%%%%%%%%%%%%%%%%%%%%%%%%%%%%%%%%
%%%%%%%%%%%%%%%%%%%%%%%%%%%%%%%%%%%%%%%%%%%%%%%%
\begin{eqnarray}
\label{eq-H2-example}
\hat{H} &=& \lambda_0 + \lambda_1 \hat{Z}_1 +
\lambda_2 \hat{Z}_2 +
\lambda_3 \hat{Z}_1 \otimes  \hat{Z}_2   \nonumber \\
&+& \lambda_4 \hat{X}_1 \otimes  \hat{X}_2 +
\lambda_5 \hat{Y}_1 \otimes  \hat{Y}_2,
\end{eqnarray}
where the coefficients  $\lambda_j = \lambda_j(R)$ are functions of the inter-nuclear distance $R$. The values of the coefficients  are obtained from  the supplementary material of Ref.~\cite{JIC18}. In Fig.~\ref{fig-fig0} the theoretically  determined (i.e. direct diagonalization of $\hat{H}$) energy spectrum is shown, and the results of the iterative excited-state extraction process are overlaid.\par

%%%%%%%%%%%%%%%%%%%%%%%%%%%%%%%%%%%%%%%%%%%%%%%%
%%%%%%%%%%%%%%%%%%%%%%%%%%%%%%%%%%%%%%%%%%%%%%%%
%%%%%%%%%%%%%%%%%%%%%%%%%%%%%%%%%%%%%%%%%%%%%%%%
%%%%%%%%%%%%%%%%%%%%%%%%%%%%%%%%%%%%%%%%%%%%%%%%
%%%%%%%%%%%%%%%%%%%%%%%%%%%%%%%%%%%%%%%%%%%%%%%%
%%%%%%%%%%%%%%%%%%
%%%%%%%%%%%%%%%%%%
%%%%%%%%%%%%%%%%%%      Fig 1-2
%%%%%%%%%%%%%%%%%%
%%%%%%%%%%%%%%%%%%%%%%%%%%%%%%%%%%%%%%%%%%%%%%%%
%%%%%%%%%%%%%%%%%%%%%%%%%%%%%%%%%%%%%%%%%%%%%%%%
%%%%%%%%%%%%%%%%%%%%%%%%%%%%%%%%%%%%%%%%%%%%%%%%
%%%%%%%%%%%%%%%%%%%%%%%%%%%%%%%%%%%%%%%%%%%%%%%%
%%%%%%%%%%%%%%%%%%%%%%%%%%%%%%%%%%%%%%%%%%%%%%%%

It must be emphasized that the ground-state in the numerical calculation is not obtained through a variational approach. Instead, the exact ground-state from diagonalization is used to construct the effective Hamiltonian at every iteration. In order to consider deviation from the exact ground-state toward a state that is the output of a VQE algorithm, a noisy ground-state is also considered. The results indicate minor differences and are shown in Fig. \ref{fig-noisy}. The details of this calculations are provided in the Supplemental Material \cite{SupplMat1}. \par

\begin{figure*}[!ht]
    \centering
    \includegraphics[width=\textwidth]{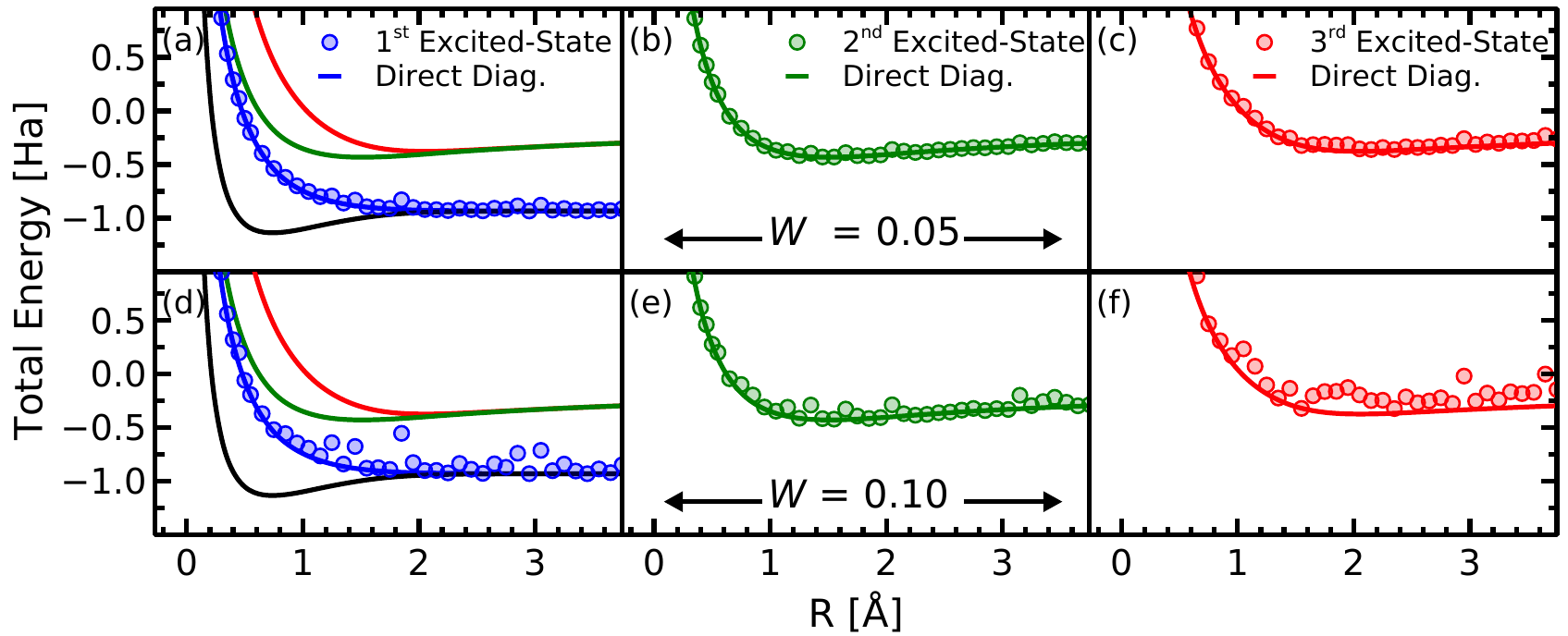}
    \caption{(Color online) The H$_2$ binding curves but under a noise model that is intended to mimic the behavior with actual HQE and VQE algorithms. The obtained excited-state energies in (a-c)  and (d-f) are for 5\% and 10\% noise models, respectively.}
    \label{fig-noisy}
\end{figure*}

\emph{Justification of the covariance assertion.---} Conceptually speaking, in statistical field calculations of thermodynamic properties, the renormalization group (RG) approach  removes some high-energy degrees of freedom from a many-body system. In this process, and as a result of the interaction between the low and high energy physics of the system, an effective Hamiltonian for the low energy physics of the problem emerges. Further assertion of self-similarity in the system close to a phase transition, connects the  parameters in the effective Hamiltonian at each RG step to the previous step by the laws of scaling \cite{MK07,RS94}.

The \emph{covariance assertion} proposed in this letter is in a sense a removal of some degrees of freedom from the problem. In contrast to RG process however, most low-lying degrees of freedom are removed. In comparison to the rescaling step of RG, a covariant form is asserted on the projected Hamiltonian and an effective one is obtained.\par

In a general sense, the RG procedure allows to partition the universe of all Hamiltonians, specified by the parameters of the Hamiltonian, onto a set of  universal classes. With this picture in mind, the \emph{covariance assertion} presumes that the problem Hamiltonian and the projected Hamiltonian belong to the same class. \par

We now discuss a class of scenarios where the \emph{covariance assertion} is justified. Consider a Hamiltonian stated as Eq~(\ref{eq-1}), but with the restriction that the $\hat{h}_j$ string operators  commute with one another, and are part of a larger set of string operators $\{\hat{h}_j\}$ that form a finite group in the mathematical sense:  The unit string operator $\hat{I}$ is included, each element has an inverse, and the set is closed under multiplication \cite{joshi1997elements}. Let us denote a set of generators of this group as $\{\hat{h}_g\}$. In addition, assume the ground-state of this Hamiltonian is well approximated by a state-vector that stabilizes the generators of this group, then the projection operator corresponding to this ground-state can be replaced  \cite{SupplMat0} by
%%%%%%%%%%%%%%%%%%%%%%%%%%%%%%%%%%%%%%%%%%%%%%%%
%%%%%%%%%%%%%%%%%%%%%%%%%%%%%%%%%%%%%%%%%%%%%%%%
%%%%%%%%%%%%%%%%%%%%%%%%%%%%%%%%%%%%%%%%%%%%%%%%
%%%%%%%%%%%%%%%%%%%%%%%%%%%%%%%%%%%%%%%%%%%%%%%%
%%%%%%%%%%%%%%%%%%%%%%%%%%%%%%%%%%%%%%%%%%%%%%%%
%%%%%%%%%%%%%%%%%%
%%%%%%%%%%%%%%%%%%
%%%%%%%%%%%%%%%%%%      Eq justify
%%%%%%%%%%%%%%%%%%
%%%%%%%%%%%%%%%%%%%%%%%%%%%%%%%%%%%%%%%%%%%%%%%%
%%%%%%%%%%%%%%%%%%%%%%%%%%%%%%%%%%%%%%%%%%%%%%%%
%%%%%%%%%%%%%%%%%%%%%%%%%%%%%%%%%%%%%%%%%%%%%%%%
%%%%%%%%%%%%%%%%%%%%%%%%%%%%%%%%%%%%%%%%%%%%%%%%
%%%%%%%%%%%%%%%%%%%%%%%%%%%%%%%%%%%%%%%%%%%%%%%%
\begin{eqnarray}
\hat{P} = \frac{1}{2^{N_g} } \prod_g (\hat{I} - \hat{h}_g).
\label{eq-Pg-ab}
\end{eqnarray}
For a Hamiltonian with the above conditions, using  Eqs. (\ref{eq-3}) and (\ref{eq-Pg-ab}), we find:
\begin{eqnarray}
\sum_j \lambda^{\prime}_j  \hat{h}_j = \sum_j \lambda_j \hat{h}_j - \frac{E_g}{{2^{N_g} } } \prod_{g=1}^{N_g} (\hat{I}- \hat{h}_g).
\label{eq-cov-con}
\end{eqnarray}
Thus, the assertion of covariance is satisfied since the products of $\hat{h}_g$ operators on the right-hand-side are contained in the set $\{\hat{h}_j\}$. In other words, if a Hamiltonian is a weighted ($\{\lambda_j\}$) sum over all elements of a (abelian) group of string operators, and the ground-state can be written as a state-vector that stabilizes the generators of this group \footnote{The ground-state is a simultaneous eigenstate of all generators; that is the justification relies on the existence of a set of symmetry operations.}, the \emph{covariance assertion} is justified.  \par

 A deviation from this simplified class of Hamiltonians occurs when, in addition to the set $\{\hat{h}_j\}$,  a few string operators that do not commute with the rest are included in the Hamiltonian. We treat this case as a perturbation.  \par

 As an example, consider the set of all the string operators on $N$ qubits with only $\hat{I}$ and $\hat{Z}$ single-qubit operators,
\begin{eqnarray}
C_z = \{\hat{I}, \hat{Z}\}^{\otimes N}.
\label{eq-Cz}
\end{eqnarray}
Trivially $C_z$ is a group.  Let
\begin{eqnarray}
\hat{H} = \sum_z \lambda_z  \hat{h}_z,
\label{eq-trivial-1}
\end{eqnarray}
where $\hat{h}_z \in C_z$, the sum is over all the elements in $C_z$,  and $\lambda_z$ are random real coefficients. For example $\lambda_z \in \left[-J, +J \right]$ and is a uniform distribution  with $J=1$.  Straightforwardly, elements in $C_z$ and thus the string operators in $\hat{H}$ commute with each others. Since the ground-state is a classical spin configuration, it can be regarded as a state that stabilizes a set of generators of $C_z$ as in Eq. (\ref{eq-Pg-ab}), and therefore Eq. (\ref{eq-cov-con}) holds. \par

 In analogy to RG, one can consider the trivial example of Eq. (\ref{eq-trivial-1}) as a fixed point in the space of parametric Hamiltonians. Then by pertubation, adding terms that do not commute with the set $C_z$ should be tolerated and the covariance assertion remains a good approximation to the excited-states. This is particularly the case in the thermodynamic limit of $N \rightarrow \infty $ \cite{SupplMat3}. \par

%%%%%%%%%%%%%%%%%%%%%%%%%%%%%%%%%%%%%%%%%%%%%%%%
%%%%%%%%%%%%%%%%%%%%%%%%%%%%%%%%%%%%%%%%%%%%%%%%
%%%%%%%%%%%%%%%%%%%%%%%%%%%%%%%%%%%%%%%%%%%%%%%%
%%%%%%%%%%%%%%%%%%%%%%%%%%%%%%%%%%%%%%%%%%%%%%%%
%%%%%%%%%%%%%%%%%%%%%%%%%%%%%%%%%%%%%%%%%%%%%%%%
%%%%%%%%%%%%%%%%%%
%%%%%%%%%%%%%%%%%%
%%%%%%%%%%%%%%%%%%      Eq C_z perturbed
%%%%%%%%%%%%%%%%%%
%%%%%%%%%%%%%%%%%%%%%%%%%%%%%%%%%%%%%%%%%%%%%%%%
%%%%%%%%%%%%%%%%%%%%%%%%%%%%%%%%%%%%%%%%%%%%%%%%
%%%%%%%%%%%%%%%%%%%%%%%%%%%%%%%%%%%%%%%%%%%%%%%%
%%%%%%%%%%%%%%%%%%%%%%%%%%%%%%%%%%%%%%%%%%%%%%%%
%%%%%%%%%%%%%%%%%%%%%%%%%%%%%%%%%%%%%%%%%%%%%%%%
Consider a perturbed version of the $C_z$ class Hamiltonian such as:
\begin{eqnarray}
\hat{H} = \sum_z \lambda_z \hat{h}_z + \sum_k \beta_k \hat{l}_k;
\label{eq-Hclass2-perturbed}
\end{eqnarray}
here $\hat{h}_z$ belongs to the set $C_z$, and $\hat{l}_k$ is a string operator such that
\begin{eqnarray}
[\hat{h}_z, \hat{l}_k] \ne 0,
\end{eqnarray}
for at least one $z$. Intuitively, the number of terms in  $\{\hat{l}_k\}$ and the average strength  of coefficients $\{ \beta_k \}$, compared to the number of operators in the set $\{\hat{h}_z\}$ and coefficients $\{ \lambda_z \} $, should be an indicator of how close we are to the class of Hamiltonians in Eq.~(\ref{eq-trivial-1}). To explore this intuitive expectation, we numerically consider many Hamiltonians in the form of Eq.~(\ref{eq-Hclass2-perturbed}) for which the following condition holds:
\begin{eqnarray}
\vert \{\hat{h}_z\}\vert > \vert \{ \hat{l}_k\} \vert,
\label{eq-con}
\end{eqnarray}
here $\vert \cdot \vert $ stands for the number of terms in the set, and  the coefficients $\lambda_z$ are chosen from a uniform distribution $\left[-1, +1\right]$. In these examples, the coefficients $\{\beta_k\}$  are considered equal to a  constant $\beta$ and $\hat{l}_k$ string operators are some tensor-product of  either $\hat{X}$ or $\hat{Y}$. We observe that for almost all values of  $\beta$, the first excited-state and excited-energy are predicted with excellent accuracy within the \emph{covariance assertion} (which is further highlighted in Supplemental Material \cite{SupplMat2}). \par

The caveat of the $C_z$  class seems to be that the number of terms  is an exponential function of the number of qubits ($2^N$). However, every element in the group can be obtained by some product of the generators. Therefore, measurement of $N$ string operator of the generators should be enough to have the needed coefficients to update the Hamilotinan at every iteration. In the context of chemistry, H$_2$ and LiH molecules appear to belong to the  $C_z$ class, based on the numerical results. \par

\emph{H$_2$.--}  In Eq.~(\ref{eq-H2-example}) the first four terms have string operators that cover all the elements in $C_z $ with $N=2$. The last two terms in Eq.~(\ref{eq-H2-example}) should be considered as perturbations. Thus, the hydrogen molecule is an example of Hamiltonians in the form of Eq. (\ref{eq-Hclass2-perturbed}). \par

Notice that an alternative is to consider the two terms $\langle Z_1\otimes Z_2,  X_1\otimes X_2 \rangle$ as the generator of an abelian group. The elements of this group are entirely in the H$_2$ Hamiltonian as well. A corresponding stabilizing-state in this case is an entangled one. However, the coefficients of the terms corresponding to this set of elements were smaller in magnitude in the entire range of inter-nuclear distance $R$, and thus we consider the H$_2$ problem to be an example of a perturbed-$C_z$. \par

\emph{LiH.--} To test the \emph{covariance assertion} method for a different fermionic quantum chemistry problem, we obtain the string of operators and corresponding coefficients of the LiH Hamiltonian from the supplemental material of Ref. \cite{AK17} where the coefficients are only at the equilibrium binding distance. In this table the string operators are presented in sets and operators in each set commute. We identify that the first set of these operators belong to $C_z$, while the only missing element in the set was $\hat{I}$. By adding this missing element,  the first round of iterations in our method is able to find the first excited-energy and excited-state ($\approx 97\%$ fidelity). In the second round of iteration, the second excited-energy is achieved with high precision, but the corresponding excited-state (eigenvector) has almost zero overlap (fidelity) with eigenvector from direct digonalization, failing to follow covariance assertion beyond the first excited-state within the $C_z$ class assumption. Discussion and results in LiH calculation are detailed in the Supplemental Material \cite{SupplMat3}. \par

\emph{Conclusion.---}
In summary, we propose a method that allows to extend
current quantum computations of the ground-state with HQC and variational algorithm to the excited-states of a given Hamiltonian. In contrast to current approaches that are shown to be successful in evaluation of the energy spectra \cite{energySpectra19}, our method determines the excited-state vector with high fidelity, given the underlying covariance assertion is closely satisfied. The success is demonstrated for H$_2$. We have also confirmed the applicability of it for LiH to some extent. \par

We should emphasize that
another advantage of our method is when the problem has a small number of qubits and measurement of all string operators is not computationally exhausting. In that case, the ground-state of the \emph{projected Hamiltonian} at every iteration can be measured rather than the effective Hamltonian. This state coincides with an excited state of the problem Hamiltonian without approximation; that is, working with the projected Hamiltonian, one can determine \emph{all} the excited state of the problem.\par

The method presented in this letter has broad impact on the use of NISQ devices. Our investigations \cite{SupplMat2} show that this method is robust to noise. Developing these tools is critical if one wishes to explore fundamental chemical processes and interpret  spectroscopic phenomena of molecular systems using gate-based quantum computation. In chemistry, for example, calculation of excited-states can lead to improved accuracy in simulating chemical reactions and stimulated processes \cite{RB07}. \par

\emph{Acknowledgements.--} This material is based upon work supported by General Atomics internal R\&D funding. We thank Eduardo Mucciolo for constructive comments. \par

\bibliographystyle{apsrev}
\bibliography{references}

\widetext

\section{Supplemental Material}
\beginsupplement
\section{S1: Steps Involved in the Method}
\label{sec:Theory}

Here we provide additional details and discussion in support of the developed method within the main manuscript.

\subsection{Projected Hamiltonian}
In the following, the eigenenergies (ground-state energy, etc.) of the problem Hamiltonian are denoted as $E_g \le E_1 \le E_2 \le \cdots $, and the superscript is dropped. The ground-state of the projected Hamiltonian,
\begin{eqnarray}
\hat{H}^{[1]} &=& (\hat{I} - \hat{P}^{[0]} ) \hat{H}^{[0]} (\hat{I}-\hat{P}^{[0]}) \nonumber \\
&=& \hat{H}^{[0]} - E_g \, \hat{P}^{[0]}
\label{eq-S3}
\end{eqnarray}
of the first iteration coincides with the first excited-state of the problem Hamiltonian $\hat{H}^{[0]}$. To see this, notice that  Eq. (\ref{eq-S3}) is independent of any representation. Spanned  by a basis set in which $\hat{H}^{[0]}$ is diagonal, we then obtain
\begin{eqnarray}
 \hat{H}^{[1]} &=& 0 \vert G \rangle \langle G \vert  + \sum_{e\ne G} E_e \vert e \rangle \langle e \vert.
\label{eq-S3-1}
\end{eqnarray}
 In Eqs.~(\ref{eq-S3}) and (\ref{eq-S3-1}), the ground-state energy $E_g$ cancels, thus, the corresponding eigenvalue of $\vert G \rangle$ in $\hat{H}^{[1]}$ is zero. \par

Upon executing  a variational algorithm to obtain the ground-state energy of $\hat{H}^{[1]}$, two possible scenarios arise: (1) $E_1 \ge 0$. In this case, the output of the  algorithm  yields $\vert G \rangle$ with eigenvalue zero, rather than the desired $E_1$ and the state $\vert e=1 \rangle$. (2) $E_1<0$. The output is the desired $\vert e=1 \rangle$ with eigenvalue $E_1$. To avoid scenario (1), prior to the out-projection, Eq. (\ref{eq-S3}), a shift to the energy spectrum must be applied; that is, add a constant to the problem Hamiltonian.   \par

The challenge is that the amount of the energy shift is nontrivial. For a generic problem, there is \emph{a priori} no information about the value of the eigenenergies, whether they are positive or negative. As long as one works with the \emph{projected Hamiltonian}, like $H^{[1]}$ above, and one measures all the string operators in the problem to construct it, the amount of shift can be any arbitrary value. \par

The complication arises when the covariance assertion is imposed.  In this case, too much energy shift may give artificially more weight to one set of string-operators than the others. This challenge is not addressed in the main text and we can only approach it case by case. However, a generic and practically useful constraint is the fact that: $E_{g} \le E_{1} \le E_{2} \le \cdots $. It can be used to design a classical algorithm that optimizes the value of the shift. More details are provided in the case of LiH. \par

\subsection{Covariance  assertion}
The idea here is to find a more intuitive reresentation of the projection operator that can justify the covariance assertion. Consider, one could write the projection operator
\begin{eqnarray}
\hat{P}&=&
\vert G \rangle \langle G \vert,
\label{eq-S2}
\end{eqnarray}
where $\vert G \rangle$ is the ground-state of the problem Hamiltonian $\hat{H}$, as
\begin{eqnarray}
\hat{P} = \frac{1}{2^{N_g} } \prod_{g=1}^{N_g} (\hat{I} - \hat{h}_g).
\label{eq-S-Pg-ab}
\end{eqnarray}
Here $\{\hat{h}_g\}$ is a set of commuting $N_g$ independent  operators, and the eigenvalue of $\hat{h}_g$ can be $\pm 1$. Since they are independent, if $N_g$ coincides with the number of qubits $N$, the state-vector that stabilizes all the operators in the set $\{\hat{h}_g\}$ has inevitably  the same dimension as $\vert G\rangle$. Physically, the existence of such a set is equivalent to existence of a set of symmetry operations that  simultaneously commute with the Hamiltonian. Thus, each eigenstate of the problem is labeled by a set of $N$ quantum numbers. \par

Clearly, for a generic given Hamiltonian, stated as a sum over some string operators, a set of string operators $\{\hat{h}_g\}$ that its elements commute with the Hamiltonian may not exist. The idea here is that the true projection operator corresponding to the ground-state, Eq.~(\ref{eq-S2}), can be approximated by the stabilizing state of a set of symmetry operators, Eq.~(\ref{eq-S-Pg-ab}).\par

Notice that if the dimension of the subspace of $\hat{P}$ in Eq.~(\ref{eq-S-Pg-ab}) is larger than $\vert G \rangle \langle G \vert$, the ground-state may be contained in $\hat{P}$, but it does not yield to Eq. (\ref{eq-S3}); the out-projection with $\hat{P}$ may remove components of the matrix that are associated to the excited-states as well. In the main text, we do not impose the condition $N_g=N$; we simply assume that the projection operator can be equivalently replaced by the projection to the stabilizing subspace of a set of commuting independent operators without further assumptions. \par

The next condition is to have all products of $\{\hat{h}_g\}$ be included in the terms $\{\hat{h}_j\}$ of the Hamiltonian, $\hat{H} = \sum_j \lambda_j \hat{h}_j$. This yields to
\begin{eqnarray}
\sum_j \lambda^{\prime}_j  \hat{h}_j = \sum_j \lambda_j \hat{h}_j - \frac{E_g}{{2^{N_g} } } \prod_g (\hat{I}- \hat{h}_g),
\label{eq-S-cov-con}
\end{eqnarray}
which means the \emph{covariance assertion} holds. Since the set of all products of $\{\hat{h}_g\}$ forms a group (in the mathematical sense) the Hamiltonian is thus a
weighted sum over the elements of this group. \par

In summary, the assertion of covariance on a Hamiltonian is justified if the following two assumptions are satisfied: 1) the Hamiltonian can be written as a weighted sum over \emph{all} elements of a finite Abelian group, and 2) the ground-state can be approximated as the state-vector that stabilizes all the commuting generators $\{\hat{h}_g\}$ of this finite group. Each generator $\hat{h}_g$ has eigenvalues $\pm1$.  Without loss of generality, suppose that the ground-state stabilizes each generator with eigenvalue $-1$, then the ground-state projection operator can be written as Eq. (\ref{eq-S-Pg-ab}). The  action of $\hat{P}$ on any state that does not belong to the ground-state subspace is zero. \par

\subsection{\texorpdfstring{$C_z$}{Cz} example}
Considering $N$ number of qubits and
\begin{eqnarray}
C_z = \{\hat{I}, \hat{Z}\}^{\otimes N},
\label{eq-S-Cz}
\end{eqnarray}
a set of possible generators is $\{\hat{Z}_1, \cdots, \hat{Z}_N\}$. Assume 1) a Hamiltonian is a weighted sum of elements in $C_z$, and 2) the ground-state lies in the subspace projected by the projection operator
\begin{eqnarray}
\hat{P} = \frac{1}{2^{N} } \prod_{i=1}^{N} (\hat{I} - (-1)^{s_i} \hat{Z}_i),
\label{eq-S-Pg-ab-1}
\end{eqnarray}
where $s_i\in \{0, 1\}$ is chosen to ensure the desired eigenvalue for $\hat{Z}_i$. After expansion, the right hand side of Eq. (\ref{eq-S-Pg-ab-1}) is a weighted sum over elements of $C_z$, which satisfies Eq. (\ref{eq-S-cov-con}). \par

\section{S2: Numerical Details}
\label{sec:noise}

Here we provide the numerical details for excited-state calculations of the hydrogen molecule. As mentioned in the main manuscript, the calculation is a classical one and neither a quantum computer nor a quantum simulator is used. The ground-state in each iteration is not a parametric ansatz. Instead, the ground-state at each iteration is obtained by diagonalizing the Hamiltonian matrix. It is expected that a variational approach, such as VQE on a HQC device, obtains an anzatz that is  close to the ground-state obtained here by diagonalization. \par

\subsection{Perfect condition}

The two-qubit Hamiltonian of the hydrogen molecule is given by
\begin{eqnarray}
\label{SupplMat:eq-H2-example}
\hat{H} &=& \lambda_0 + \lambda_1 \hat{Z}_1 +
\lambda_2 \hat{Z}_2 +
\lambda_3 \hat{Z}_1 \otimes  \hat{Z}_2   \nonumber \\
&+& \lambda_4 \hat{X}_1 \otimes  \hat{X}_2 +
\lambda_5 \hat{Y}_1 \otimes  \hat{Y}_2,
\end{eqnarray}
as shown in the main text. The coefficients  $\lambda_j = \lambda_j(R)$ are a function of the inter-nuclear distance $R$. These coefficients are obtained from the supplementary materials in Ref.~\cite{JIC18}.\par

 At a given $R$, we insert the values of $\{ \lambda_j(R) \}$ in the Hamiltonian, Eq. (\ref{SupplMat:eq-H2-example}), and diagonalize the Hamiltonian and obtain the ground-state vector. This step  is equivalent to (an error-free) optimization of the parametric ansatz in the VQE method for quantum hardware. Next, we take the expectation value of each of the six $\hat{h}_j$ terms in  Eq. (\ref{SupplMat:eq-H2-example}) and denote with $f_j$. Using $f_j$ value, $\lambda_j$ is updated and a new Hamiltonian is constructed,
 \begin{eqnarray}
\hat{H}^{[i]} = \sum_j^{N_h} \lambda^{[i]}_j \hat{h}_j \rightarrow \hat{H}^{[i+1]} = \sum_j^{N_h} \lambda^{[i+1]}_j \hat{h}_j,
\label{eq-S-5}
\end{eqnarray}
where the coefficients are updated by
\begin{eqnarray}
\lambda^{[i+1]}_j  = \lambda_j^{[i]} - E_g^{[i]}\,\,f^{[i]}_j,
\label{SupplMat:eq-renorm}
\end{eqnarray}
with $i=0$. This is the end of first iteration under covariance assertion. Now, the ground-state/energy of $H^{[1]}$, which we obtain by exact digonalization,  is the first excited-state/energy of $H^{[0]}$. The next iteration proceeds by repeating the steps above using the updated effective Hamiltonian. The iterations can continue and therefore excited-states can be determined sequentially. \par

\subsection{Including noise and mimicking a variational algorithm}

To simulate the imperfect VQE procedure, we assume a noise model as follows: At every iteration a set of random real numbers $\{ \epsilon \vert \epsilon \in \left[-W, W\right] \}$ are considered. $W$ represents the maximum strength of the error. Next, the entries of the state-vector $\vert \psi \rangle = \sum_j c_j \vert j \rangle $ from the diagonalization, expressed in some computational basis set $\{\vert j \rangle \}$, are corrupted: $c_j \rightarrow c_j +\epsilon_j$. This replaces the state-vector from exact diagonalization (perfect condition). This is then  used to evaluate $\{f_j\}$ coefficients, and other steps as before (Eqs. (\ref{eq-S-5}) and (\ref{SupplMat:eq-renorm})).

Notice that the considered noise model does not preserve the wave-function norm. This is a realistic assumption, as the choice of an ansatz for VQE may be a poor approximation of the exact wave-function, and the experimental quantum gates performing the corresponding quantum circuit may also deviate from executing unitary operators. \par

\begin{figure*}[ht]
    \centering
    \includegraphics[scale=1]{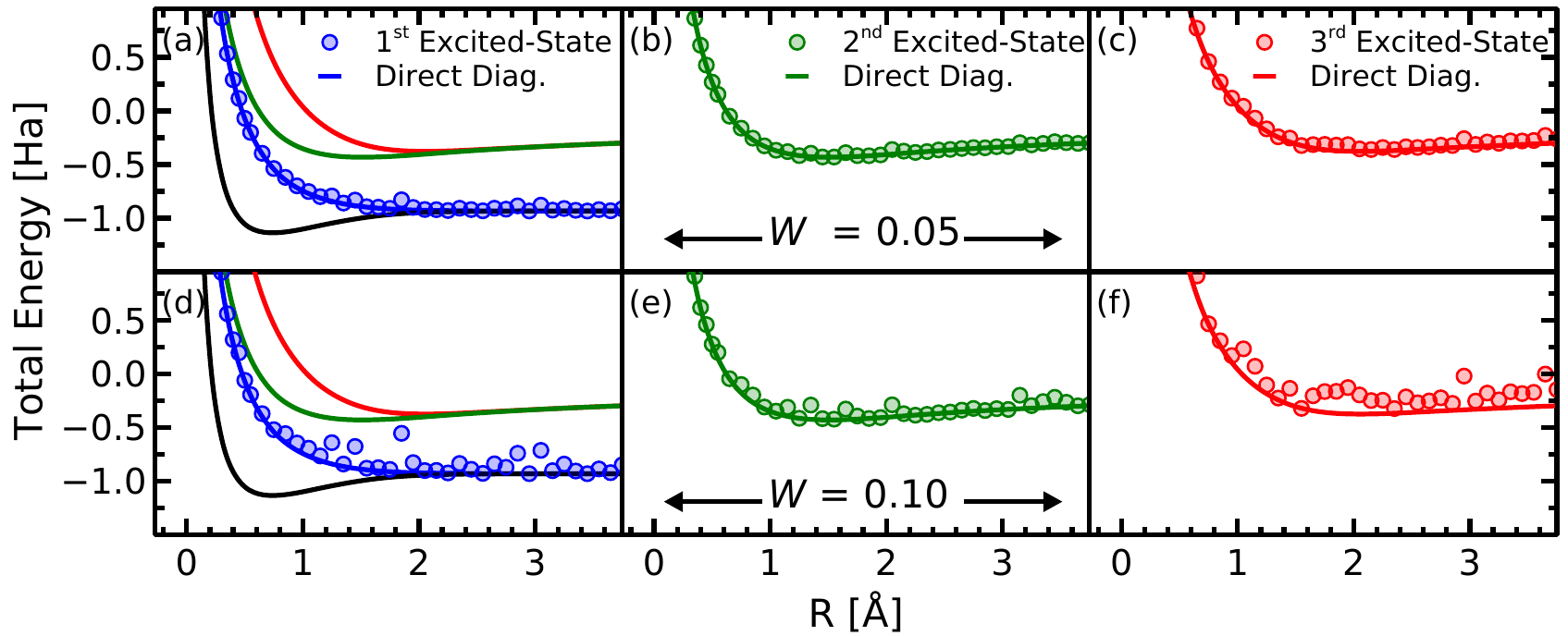}
    \caption{(Color online) Calculated excited-states for H$_2$ at different inter-nuclear distances when a noise model is considered. The noise is used to create a corrupt version of the exact states used in our calculation.  The top plots are for a maximum noise factor of $W = 0.05$ and bottom plots for $W = 0.10$.}
    \label{SupplMat:noise}
\end{figure*}

In the presence of the noise, we find that the resultant excited-state energies remains close to the unperturbed values as shown in Fig.~\ref{SupplMat:noise}. The fidelity of the states remained  above $95\%$ for most of the $R$ points. The fidelity decreases when moving toward higher excited states, which
suggests that our proposed method is robust in the presence of a low-amount of  noise/imperfection and therefore is suitable for the determination of the excited-states on quantum hardware.  We note, that in the presence of noise it may be difficult to obtain chemical accuracy (e.g., $\sim$1 milli-Hartree) typically desired in quantum chemistry calculations. \par

\section{S3: Perturbed  \texorpdfstring{$C_z$}{Cz}}
\label{sec:excited}

In this part, we focus on describing the Hamiltonians that fall under  the characteristic form,
\begin{eqnarray}
\hat{H} = \sum_z \lambda_z \hat{h}_z + \sum_k \beta_k \hat{l}_k.
\label{eq-S-Hclass2-perturbed}
\end{eqnarray}
Here $\hat{h}_z$ belongs to the set $C_z$ (Eq.~(\ref{eq-S-Cz})), and $\hat{l}_k$ is a string operator such that
\begin{eqnarray}
[\hat{h}_z, \hat{l}_k] \ne 0
\end{eqnarray}
for at least one $z$. In particular, we choose  the following prototype,
\begin{eqnarray}
\hat{H} = \sum_{z} \lambda_z \hat{h}_z + \hat{V},
\label{eq-H-num}
\end{eqnarray}
where the sum in the first term on the right-hand side is over the elements in $\hat{h}_z \in C_z$, and the corresponding $\lambda_z $ is drawn from  the uniform distribution $\lambda_z \in \left[ -1, +1 \right]$. \par

For example, consider the perturbation $\hat{V}$ to have the following prototype,
\begin{eqnarray}
V = \sum_x \beta_1 \hat{l}_x + \sum_y \beta_2 \hat{l}_y,
\end{eqnarray}
where $\beta_1$ and $\beta_2$ are considered constants, and the string operators $\hat{l}_x$ are
\begin{eqnarray}
\hat{l}_x \in \{ \hat{X}_1, \cdots, \hat{X}_N,
\,\, \hat{X}_1\hat{X}_2, \cdots, \hat{X}_{N-1}\hat{X}_N \}, \nonumber \\
\end{eqnarray}
and the set of of $\{\hat{l}_y\}$ string operators is obtained by replacing $\hat{X}$  with $\hat{Y}$ in every $\hat{l}_x$. In the following we consider  $\beta_1 \in \left[ 0.1, 0.3, 0.5, 0.7, 0.9,1.1\right]$ and $\beta_2=0$. Similar results  are checked to be true for $\beta_1 = \beta_2$. \par

Now, consider a set of random coefficients $\{ \lambda_z \}$. After replacing for the coefficients in Eq. (\ref{eq-H-num}), the Hamiltonian is diagonalized and the ground-state is used to evaluate steps involved in the first round of iteration of the method. We have performed these steps for different system sizes and measured the energy differences between the exact and predicted first excited state energy, as well as the fidelty of the predicted first excited-state vector with respect to the exact one.  \par

\begin{figure}[!ht]
    \includegraphics[width=8.6cm]{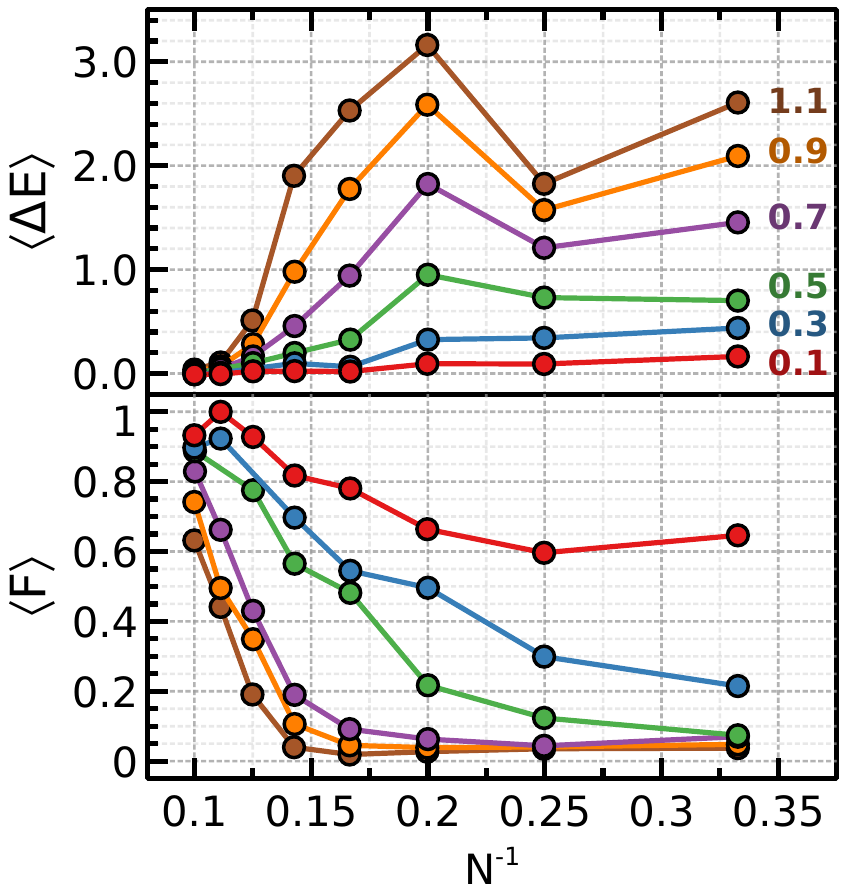}
    \caption{(Color online) Thermodynamic limit of energy difference per exact value (top) and the fidelity (bottom) for the class of Hamiltonians stated as perturbed-$C_z$, for various $\beta_1$ and $\beta_2 = 0$.  The trend exhibits the convergence at the thermodynamic limit.}
    \label{fig-quality}
\end{figure}

Our purpose is to demonstrate improvement of the covariance assertion in the thermodynamic limit of $N \rightarrow \infty$. For measuring deviation from the exact eigenvalue, we define
\begin{eqnarray}
\Delta E = \frac{\vert E_{pred} - E_{exact}\vert }{\vert E_{exact} \vert },
\label{eq-dE}
\end{eqnarray}
where $\vert \cdot \vert $ stands for absolute value. $E_{exact}$ ($E_{pred}$) is the exact (predicted) energy, and the fidelity
\begin{eqnarray}
F = \vert \langle e_1^{pred} \vert e^{exact}_1 \rangle \vert,
\label{eq-F}
\end{eqnarray}
where the $\vert e^{exact}_1 \rangle $ ($\vert e^{pred}_1 \rangle$) is the exact (predicted) first-exited state. These quantities are shown in Fig. \ref{fig-quality} as a function of inverse of $N$. For smoothness, at a given system size (number of qubits) $N$, $F$ and $ \Delta E $ are averaged over thirty set of random  $\{\lambda_j\}$ realizations  in Eq.~ (\ref{eq-H-num}).  As the thermodynamic limit  $N\rightarrow \infty$ is approached, $F\rightarrow 1$ and $\Delta E  \rightarrow 0$, as can be seen in the figure. \par

\section{S4: Related Example - Lithium Hydride Excited States}
\label{sec:LiH2}

Another problem that we identify as closely belonging to the class of Hamiltonians in Eq. (\ref{eq-H-num}), is the  LiH molecule, which is an example of the perturbed-$C_z$ class of Hamiltonians. \par

The ground-state energy of this molecule has been experimentally evaluated on quantum hardware via the VQE approach and the Jordan-Wigner transformed fermionic Hamiltonian of the LiH molecule \cite{AK17}. In this work, we use the table of string operators $\{\hat{h}_j \}$ and their corresponding coefficients $\{\lambda_j \}$ at the equilibrium bond distance of LiH as stated in the Supplemental Materials of ref.~\cite{AK17}. Inspection of these string operators leads to identifying the Hamiltionian as being an instance of Eq.~(\ref{eq-H-num}), where $V$ now contains all the terms in the LiH Hamiltonian (see Ref. ~\cite{AK17}) beside those already contained in the $C_z$ with $N=4$. \par

\begin{figure}[!h]
    \includegraphics[scale=1]{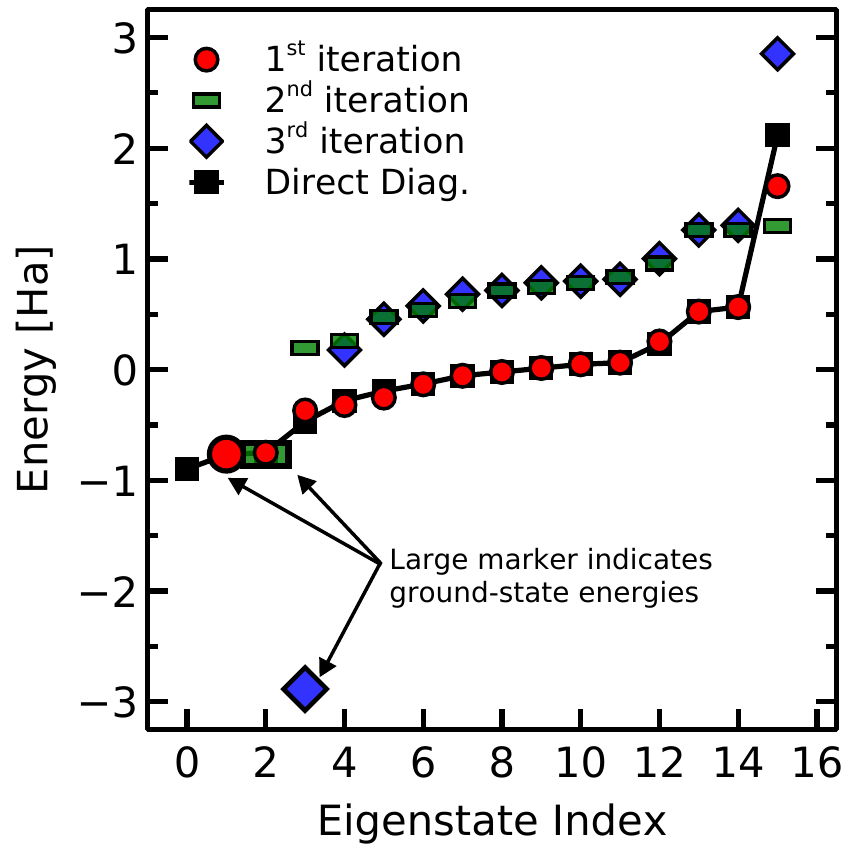}
    \caption{(Color online) The eigenvalue spectrum of the LiH molecule, at equilibrium bond distance (nuclear part of the  energy is not included). The ground-state energies of three iterations obtained using the proposed excited-state projection method are shown with larger markers. The first and second iterations have ground-state energies that match the exact values.  The third iteration shows deviation from the exact value. The smaller markers in each series represent other eigenenergies from the diagonalization of the projected Hamiltonian after covariance assertion.}
    \label{SupplMat:LiH}
\end{figure}

In light of this view, we can follow the steps of the proposed method and covariance assertion at every step. We obtain up to the third excited energy of the LiH molecule, with $97\%$ fidelity  for the first-excited state. The obtained states are compared with the exact eigenvectors from the direct diagonalization.  The eigenenergies of the effective Hamiltonians of the first three iterations are depicted along with the exact eigenvalues of the Hamiltonian in Fig.~\ref{SupplMat:LiH}. The low-lying  obtained eigenenergies $E_1 = E^{[1]}_g$, $E_2 = E^{[2]}_g$, and $E_3 = E^{[3]}_g$, are highlighted by larger marker sizes in the figure. The smaller marker sizes show the rest of the effective Hamiltonian spectrum  in that iteration. As one can see, the spectrum at the second iteration differ from the exact ones at high energies. This leads to the failure of covariance assertion in the following iteration (i.e., third iteration).\par

For this problem, we add the missing $I \in C_z$ element that is not stated in the table of the Ref~\cite{AK17}. In principle, one can add this element with any corresponding initial $\lambda_{I}$, as this value is  just a shift in the energy spectrum.  However, once the covariance assertion is applied, we observe that for different values of $\lambda_I$ the results of the method are different. To accomplish a meaningful result, we consider a range of $\lambda_I$ from $-10$ to $10$ and with small increments of $0.2$. For each possible $\lambda_I$, we calculate the first, second, and third excited energies using the \emph{covariance assertion}; that is, asserting that the Hamiltonian is a sum over complete set of $C_z$, plus the rest of the string operators in the original Hamiltonian as a perturbation. For each $\lambda_I$ value, the first-excited energy state must be larger than or equal to the ground-state energy, the second-excited state must be larger than or equal to the first-state energy, etc. Enforcing these constraints during a sweep over $\lambda_I$, we find an optimized value of $\lambda_I$. At the end of the calculation this shift of energy is subtracted. As shown in  Fig. (\ref{SupplMat:LiH}), the $E^{[3]} \ge E_g^{[2]}$ condition could not be satisfied. This indicates that the projected Hamiltonian at this iteration significantly deviates from the assumed perturbed-$C_z$ class of Hamiltonians. Therefore, the covariance assertion that the Hamiltonian is a renormalized form of a perturbed $C_z$ class is not valid for the energies above this excited-state. \par

\end{document}